\documentclass[prc,twocolumn,showpacs,preprintnumbers,amsmath,amssymb,superscriptaddress,floatfix,nofootinbib]{revtex4}
\usepackage{graphicx}
\usepackage{amsmath}
\usepackage{amsfonts}
\usepackage{slashed}
\usepackage{amssymb}
\usepackage{hyperref}
\usepackage{color}
\usepackage{ulem}
\hypersetup{breaklinks=true}
\newcommand{\sm}[1]{\scriptscriptstyle{#1}}

\newcommand{\boldm}[1]{\mbox{\boldmath{$#1$}}}
  \def\CL{{\cal L}}
\def\CM{{\cal M}}

\usepackage{xcolor}

\begin{document}

\title{Sifting out the neutral and charged components of $X(3872)$ from a spin observable}

\author{Ke Wang}
\affiliation{School of Nuclear Science and Technology, University of Chinese Academy of Sciences, Beijing 101408, China}

\author{Yu-Fei Wang}
\affiliation{School of Nuclear Science and Technology, University of Chinese Academy of Sciences, Beijing 101408, China}

\author{Bo-Chao Liu}
\affiliation{Ministry of Education Key Laboratory for Nonequilibrium Synthesis and Modulation of Condensed Matter, School of Physics, Xi'an Jiaotong University, Xi'an 710049, China}

\author{Fei Huang}
\email{huangfei@ucas.ac.cn}
\affiliation{School of Nuclear Science and Technology, University of Chinese Academy of Sciences, Beijing 101408, China}

\date{\today}

\begin{abstract}

In this work, we propose a new method to detect the composition of the $X(3872)$ state. Based on a widely accepted interpretation that $X(3872)$ is a weakly bound $S$-wave molecule of the $D^{*0} \bar{D}^{0}$ (neutral) and $D^{*+} D^{-}$ (charged) configurations, the process $X(3872) \to \bar{D}^{*0} D^0$ is described by one-loop triangle diagrams with the corresponding components as intermediate particles, and with the parameters constrained by the experimental branching ratio. Unlike the mild behavior of the neutral component, the charged component manifests itself as a distinct structure near the $D^{*+} D^{-}$ threshold at $3.880$ GeV, in the $M_{\sm{\bar{D}^{*0} D^0}}$ invariant mass distribution of the spin density matrix element $\rho^{\sm{D^*}}_{00}$. Hence the line shape of $\rho^{\sm{D^*}}_{00}$ near this structure depends sensitively on the proportions of the two configurations. We propose high-precision measurements of this matrix element by future BESIII and Belle experiments to elucidate how a molecular $X(3872)$ state is composed. 

\end{abstract}

\maketitle

\section{INTRODUCTION}
\label{introductiuon}

The study of the internal constituents of hadrons is one of the important topics for the understanding of quantum chromodynamics (QCD) in the nonperturbative regime. The conventional quark model proposed by Gell-Mann and Zweig~\cite{Gell-Mann:1964ewy,Zweig:1964ruk} categorizes the hadrons into mesons with a quark-antiquark pair and baryons with three quarks. Based on this model, a simple and beautiful description for the ground-state hadrons has been established in last century. However, in the last two decades more and more exotic hadron candidates have been observed experimentally, extending the knowledge of hadron structures to a deeper level. Many other hadron configurations that are compatible with the color confinement, such as compact states with four or more quarks/antiquarks, hadronic molecules, glueballs, and hybrids, spark the renaissance of hadron spectroscopy~\cite{Chen:2016qju,Hosaka:2016pey,Esposito:2016noz,Ali:2017jda,Guo:2017jvc,Karliner:2017qhf,Liu:2019zoy,Guo:2019twa,Brambilla:2019esw,Chen:2022asf,Kalashnikova:2018vkv}. Especially, the picture of hadronic molecules has garnered substantial interest, since many of exotic candidates are located close to some certain thresholds of two-hadron systems. Specifically, in the molecular picture, an exotic state is described as a compound of two hadrons formed by residual interactions, i.e. direct hadron exchanges. Such states are weakly bounded and thus lie rather close to the corresponding thresholds, analogous to the deuteron. As more and more states are justified as hadronic molecules, the clarification of their components becomes particularly important. 

In 2003, an exotic charmonium-like state $X(3872)$ was first observed by the Belle collaboration in the reaction $B^+ \to K^+ X(3872) \to K^+ \pi^+\pi^-J/\psi$~\cite{Belle:2003nnu}, and promptly confirmed by CDF~\cite{CDF:2003cab}, D0~\cite{D0:2004zmu}, and BaBar~\cite{BaBar:2004oro} collaborations. Subsequently, its quantum numbers were determined to be $J^{PC}=1^{++}$ by the LHCb collaboration~\cite{LHCb:2013kgk}. Although these quantum numbers indicate that $X(3872)$ seems to be a genuine $2^3P_1$ charmonium, its mass and width, as well as the non-negligible isospin-violation effect quantified by the ratio of the coupling constants $R_X = g_{X \psi \rho} / g_{X \psi \omega} \approx 0.3$~\cite{Braaten:2005ai,Suzuki:2005ha}, make its nature ambiguous~\cite{Godfrey:1985xj,BaBar:2010wfc}. Moreover, since $X(3872)$ lies extremely close to the $D^{*0} \bar{D}^{0}$ threshold and strongly couples to this channel, its interpretation as a $D^* \bar{D}$ molecule is natural and has been widely accepted. Hereafter we do not distinguish $D^{*0} \bar{D}^{0}$ from $\bar{D}^{*0} D^{0}$ following the treatment in Refs.~\cite{Belle:2008fma,Belle:2023zxm}. For a comprehensive review of $X(3872)$ in the molecular model, see Ref.~\cite{Kalashnikova:2018vkv} and the references therein. 

However, the exploration of $X(3872)$ is not limited to the explanation above. There are two configurations -- $D^{*0} \bar{D}^{0}$ (neutral) and $D^{*+} D^{-}$ (charged), the proportions of which are not clearly known due to the isospin breaking effect. Inspired by an earlier study~\cite{Dubynskiy:2007tj} on the relative rates of $X(3872)\to \pi\chi_{cJ}$ transitions, Refs.~\cite{Wang:2022qxe,Wang:2025zss} have revealed that the relative rates of $X(3872)\to VV/VP$ and $X(3872)\to \gamma V$ transitions depend sensitively on the proportion of the charged and neutral constituents ($V$, $P$, and $\gamma$ stand for the vector meson, the pseudoscalar meson, and the photon, respectively). Moreover, in Ref.~\cite{Zhang:2024fxy}, an isovector $W_{c1}$ is predicted near the $D^{*+} D^-$ threshold by chiral effective field theory. Such a state, if confirmed, gives further evidence on the molecular nature of $X(3872)$. By fitting experimental data, it has been found in Ref.~\cite{Ji:2025hjw} that both $X(3872)$ and $W_{c1}$ are molecules formed by $D^{*0} \bar{D}^0$ and $D^{*+} D^-$ coupled-channel interactions. All in all, the neutral $D^{*0} \bar{D}^0$ and the charged $D^{*+} D^-$ configurations of $X(3872)$ need, and in principle can also be, distinguished from the observables. In this paper, we propose a new method thereon by means of a spin observable based on triangle diagrams. 

The triangle diagrams are famous for its special kinematic effect -- the triangle singularity (TS) proposed by Landau in 1959~\cite{Landau:1959fi} , which plays an essential role for solving some puzzles in hadron physics nowadays~\cite{Wu:2011yx,Aceti:2012dj,Wu:2012pg,Achasov:2015uua,Du:2019idk,Jing:2019cbw,Guo:2019qcn,Sakai:2020ucu,Molina:2020kyu,Sakai:2020crh,Yan:2022eiy,Wang:2023xua,Wang:2024ewe}. In addition to the TS, the two-body threshold singularities are also kinematic effects that can induce structures in the invariant mass spectrum. For example, References~\cite{Wang:2022wdm,Wang:2023xua} have shown that kinematic effects may cause significant spin effects, establishing a connection between the underlying interaction mechanism and the experimental observation. In the molecular scenario, the one-pion exchange (OPE) potential is essential for forming the shallow bound state $X(3872)$~\cite{Meng:2022ozq,Wang:2013kva,Xu:2021vsi,Li:2012cs}. Consequently, the transition from $X(3872)$ to $\bar{D}^{*} D$ must proceed first with an $X D^* \bar{D}$ vertex, then with the OPE interaction between the $D^* \bar{D}$, namely a triangle loop. Such a diagram should exhibit observable structures near the kinematic singularities. Specifically, in this work, we investigate the transition $X(3872) \to \bar{D}^{*0} D^0$ through the spin density matrix element (SDME) $\rho^{\sm{D^*}}_{00}$ of the final state $\bar{D}^{*0}$. Especially, the charged constituents $D^{*+} D^-$ in $X(3872)$ develop a threshold singularity in the $\bar{D}^{*0} D^0$ invariant mass spectrum. In contrast to the contribution from the neutral configuration which produces no kinematic singularity, the one from the charged configuration contains a manifest peak at the $D^{*+} D^{-}$ threshold. Furthermore, it is worth noticing that such structure can hardly be seen in some other observables of the same process, such as cross sections or invariant mass distributions of the decays. In those observables, the sharp Breit-Wigner peak of $X(3872)$ would make the kinematic structure invisible, whereas the spin observable we propose here, as a ratio of differential cross sections, eliminates the Breit-Wigner peak.

This paper is organized as follows. In Sec.~\ref{formalism}, we present the theoretical framework and formulate the amplitudes for the reaction $X(3872) \to \bar{D}^{*0} D^0$. In Sec.~\ref{results}, we show the numerical results and discuss their implications. Finally, we summarize our findings and conclusions in Sec.~\ref{summary}.

\section{Formalism}
\label{formalism}

\begin{figure*}[tbp]
        \begin{center}
        \includegraphics[scale=0.5]{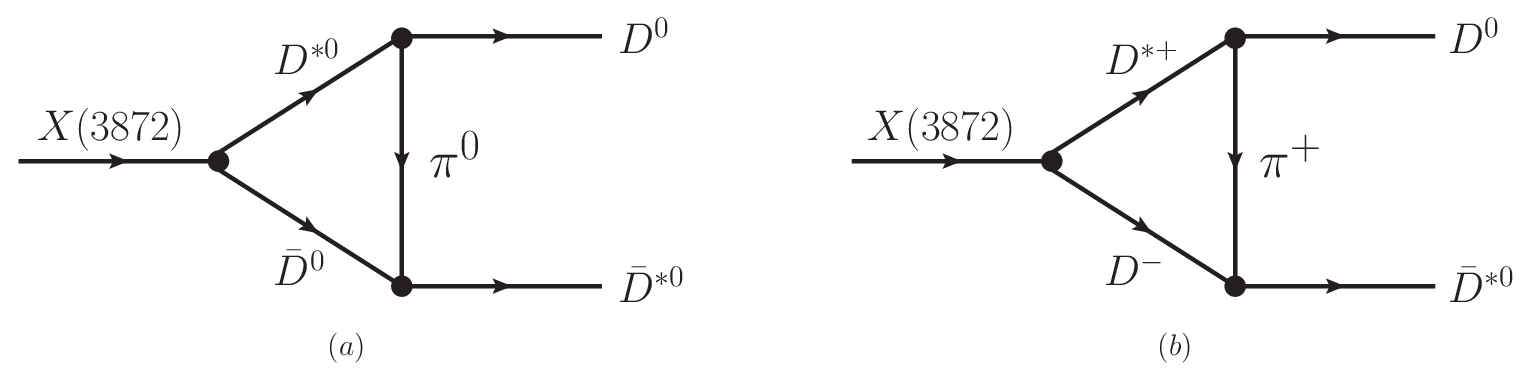}
        \caption{The Feynman diagrams for reaction $X(3872) \to \bar{D}^{*0} D^0$.} \label{Fig:Feynman}
        \end{center}
\end{figure*}

In this work, we investigate the ``decay'' process $X(3872) \to \bar{D}^{*0} D^0$, depicted in Fig.~\ref{Fig:Feynman}, within the molecular picture. Our goal is to develop a method to probe the proportions of neutral and charged configurations in $X(3872)$ through a spin observable of the final vector meson $\bar{D}^{*0}$. Note that $X(3872)$ is a shallow bound state below the $\bar{D}^{*0} D^0$ threshold. The nominal ``decay'' is actually a transition process that can be theoretically described by vertex functions (like the triangle diagrams). Experimentally it can also be measured in some physical processes containing both the transition and the production of an intermediate virtual $X(3872)$ state.

Following Refs.~\cite{Dong:2009yp,Wu:2021udi,Cai:2024glz}, we assume that $X(3872)$ is an $S$-wave molecular state arising from the superposition of $D^{*0}\bar{D}^0$ and $D^{*+}D^-$, 
\begin{equation}\label{EQ:thetadef}
\begin{split}
    \left |X(3872) \right \rangle &= \frac{\cos\theta}{\sqrt{2}} \left( \left | \bar{D}^{*0}D^0 \right \rangle + \left | D^{*0}\bar{D}^0 \right \rangle \right)\\
    &\quad + \frac{\sin\theta}{\sqrt{2}} \left( \left | D^{*-}D^+ \right \rangle + \left | D^{*+}D^- \right \rangle \right)\\
    &\equiv \cos\theta \big|\Phi_n \big\rangle+\sin\theta \big|\Phi_c \big\rangle, 
\end{split}
\end{equation}
where $\big|\Phi_n \big\rangle(\big|\Phi_c \big\rangle)$ stands for the pure neutral (charged) constituents; $\theta$ is the mixing angle describing the proportion of $\big|\Phi_n \big\rangle$ and $\big|\Phi_c \big\rangle$. In other words, $\theta=0$ and $\theta=\pi/2$ denote $X(3872)$ as a bound state composed of pure neutral and charged constituents, respectively. Besides, if $X(3872)$ is a pure isospin singlet, $\theta=\pi/4$. Then, we can describe the interaction between $X(3872)$ and its constituents with following effective Lagrangian~\cite{Wu:2021udi,Guo:2014taa,Wang:2022qxe},
\begin{eqnarray}
    \CL_{XD^*\bar{D}} &=& \frac{g_n}{\sqrt{2}} X^\mu \left( \bar{D}^{*0}_\mu D^0 + D^{*0}_\mu \bar{D}^0 \right) 
    \nonumber \\ & &
    + \frac{g_c}{\sqrt{2}} X^\mu \left( D^{*-}_\mu D^+ + D^{*+}_\mu D^- \right),\label{LXDD}
\end{eqnarray}
where $g_n$ and $g_c$ are the coupling constants of $X(3872)$ to its neutral and charged constituents, respectively. They are related to the mixing angle $\theta$: 
\begin{eqnarray}
    g_n = g^{\text{eff}}_n \cos\theta\ ,\quad g_c = g^{\text{eff}}_c \sin\theta ,
\end{eqnarray}
where $g^{\text{eff}}_{n}$ and $g^{\text{eff}}_{c}$ are the effective couplings of the molecular state $X(3872)$ to $\big|\Phi_n \big\rangle$ and $\big|\Phi_c \big\rangle$ constituents, respectively. 

Based on Weinberg's criterion on the compositeness~\cite{Weinberg:1965zz}, for an $S$-wave shallowly bound molecule, its effective coupling to the constituents depends on the compositeness $A$ and the binding energy $\epsilon$~\cite{Weinberg:1965zz,Baru:2003qq,Guo:2013zbw,Wang:2025zss,Molina:2020kyu,Wang:2022qxe}. Specifically, we take the convention in Refs.~\cite{Wang:2025zss,Molina:2020kyu,Wang:2022qxe} for $X(3872)$,
\begin{eqnarray}
    (g_i^{\text{eff}})^2 = 16\pi A_i \sigma_i^2 \sqrt{\frac{2\epsilon_i}{\mu_i}},
\end{eqnarray}
where the subscript $i=n (c)$ represents the neutral (charged) constituents, $\sigma_i$ is the corresponding threshold, e.g. $\sigma_c = m_{D^{*+}} + m_{D^-}$, and the reduced mass and the binding energy are expressed by $\mu = m_{D^*} m_{\bar{D}} / (m_{D^*} + m_{\bar{D}})$ and $\epsilon = m_{D^*} + m_{\bar{D}} - m_X$, respectively. Note that, under the specified $g_i^{\text{eff}}$'s above, the parameter $\theta$ is mathematically equivalent to the ratio $g_n/g_c$ which indicates the isospin composition of $X(3872)$. 

As mentioned in the introduction, we make no distinction between different charge-conjugation states in the process $X(3872) \to \bar{D}^{*0}D^0 + c.c.$, which is consistent with the experimental measurements of this branching ratio~\cite{Belle:2008fma,Belle:2023zxm}. Technically, ``no distinction'' is realized here by a convention that only mentions the $\bar{D}^{*0}D^0$ final state. Then for the pure charged constituents $|\Phi_c\rangle$, we only need to mention its $D^{*+}D^-$ configuration. Therefore its compositeness value is $A_c=1$ under this convention. Similarly, the pure neutral constituents $|\Phi_n\rangle$ needs only $D^{*0}\bar{D}^0$ with $A_n=1$. Taking the masses from PDG~\cite{ParticleDataGroup:2024cfk}, we have $g^{\text{eff}}_n=2.62$ GeV and $g^{\text{eff}}_c=9.94$ GeV. Note that there is another critical point at $\cos^2\theta\simeq 0.94$, leading to $g_n=g_c$, namely the isospin symmetry is effectively restored in Eq.~\eqref{LXDD}. 

We would like to note that there can be an alternative convention accounting for different charge-conjugate states explicitly: $|\Phi_c\rangle$ includes both $D^{*+}D^-$ and $D^{*-}D^+$, while $|\Phi_n\rangle$ includes both $D^0 \bar{D}^{*0}$ and $\bar{D}^0 D^{*0}$, resulting in four Feynman diagrams (with two additional diagrams for the $\bar{D}^0 D^{*0}$ final state). In this convention the compositeness values are halved since the channel space is doubled, yielding the change $g_{n,c}\to g_{n,c}/\sqrt{2}$. Meanwhile, the contribution from the extra two Feynman diagrams compensates for that change. Finally, both conventions yield the same result. In other words, our convention correctly reproduces the experimental branching ratio even if it only mentions the $D^0 \bar{D}^{*0}$ final state, as shown in Fig.~\ref{Fig:Feynman}.

For the $D^*D\pi$ vertex appearing in Fig.~\ref{Fig:Feynman}, we adopt the following general form of the effective Lagrangian,
\begin{equation}
    \CL_{D^*D\pi} = ig_{\sm{D^*D\pi}} D^{*\mu} \partial_\mu \pi \bar{D}.
\end{equation}
Every coupling constant $g_{\sm{D^*D\pi}}$ can be determined through the corresponding partial decay width using
\begin{equation}
    \Gamma_{D^* \to D\pi} = \frac{g^2_{\sm{D^*D\pi}}}{24\pi} \frac{|\boldm{p}_{\pi}|^3}{m^2_{\sm{D^*}}}, \label{Eq:D_width}
\end{equation}
where $\boldm{p}_{\pi}$ denotes the three momentum of the $\pi$ in the rest frame of $D^*$. For the charged $D^{*\pm}$ states, the width can be obtained from PDG \cite{ParticleDataGroup:2024cfk}, $\Gamma_{D^{*\pm}} = (83.4 \pm 1.8)$ keV. For $D^{*0}$, although its width only has an upper limit as $\Gamma_{D^{*0}} < 2.1$ MeV, Ref. \cite{Guo:2019qcn} gets the value according to the isospin symmetry, $\Gamma_{D^{*0}} = (55.3 \pm 1.4)$ keV. The central values of those decay widths lead to the coupling constants in Table \ref{Tab:couplings}.

With the Lagrangian densities above, we can straightforwardly write down the decay amplitudes from the triangle loop diagrams shown in Fig.~\ref{Fig:Feynman}: 
\begin{eqnarray}
    \CM_a &=& g_a \, \varepsilon^\mu_{X} \varepsilon^{*\nu}_{\bar{D}^{*0}} \int\frac{\text{d}^4 p_{\sm{\bar{D}}}}{(2\pi)^4} G^{1}_{\mu\alpha}(p_{\sm{D^{*}}})  p^\alpha_{\sm{\pi}} p_{\sm{\pi,\nu}} 
    \nonumber \\ & & 
    G^0(p_{\sm{\bar{D}^0}}) G^0(p_{\sm{\pi}}) F(p_{\sm{\pi}}) 
    \nonumber \\ &\equiv& 
    g_a \, \varepsilon^\mu_{X} \varepsilon^{*\nu}_{\bar{D}^{*0}} \CM^{\text{Loop}}_{a,\mu\nu}, \label{Amp:a} \\
    \CM_b &\equiv& g_b \, \varepsilon^\mu_{X} \varepsilon^{*\nu}_{\bar{D}^{*0}} \CM^{\text{Loop}}_{b,\mu\nu}, \label{Amp:b} \\
    \CM^{\text{Total}} &=&  \CM_a + \CM_b, \label{Amp:all}
\end{eqnarray}
where $\varepsilon_X$ and $\varepsilon^*_{\bar{D}^{*0}}$ are the spin polarization vectors of $X(3872)$ and $\bar{D}^{*0}$, respectively. The effective couplings $g_a$ and $g_b$ read
\begin{eqnarray}
    g_a &=& \frac{g_n}{\sqrt{2}} g^2_{\sm{D^{*0}D^0\pi^0}}, \\
    g_b &=& \frac{g_c}{\sqrt{2}} g_{\sm{D^{*+}D^0\pi^+}} \times \sqrt{2} g_{\sm{D^{*0}D^0\pi^0}}.
\end{eqnarray}
The $G^J$'s denote the propagators of the intermediate particles with spin $J$~\cite{Wu:2021udi}: 
\begin{eqnarray}
    G^0(p) &=& \frac{i}{p^2-m^2}\label{Eq:pp0}, \\
    G^1_{\mu\nu}(p) &=& -\frac{i \left(g_{\mu\nu}-p_\mu p_\nu/m^2\right)}{p^2-m^2+im\Gamma}\label{Eq:pp1},
\end{eqnarray}
where $q$, $m$, and $\Gamma$ are the four momentum, mass, and constant width of the intermediate particle, respectively. Meanwhile, a phenomenological form factor $F(p)$ is introduced to make the loop integral convergent and to include the off-shell effects of the intermediate $\pi$ meson 
\begin{equation} 
    F(p) = \frac{\Lambda^4}{\Lambda^4+(p^2-m_\pi^2)^2},  \label{Eq:form_factor}
\end{equation}
where $\Lambda$ is the cutoff parameter. Since the energy range we concern is rather narrow and close to the threshold, the off-shell effects of $D^*$ and $\bar{D}$ are small, and thus we can disregard their form factors. Here we do not adopt $F(p)=\left(\Lambda^2-m^2\right) / \left(\Lambda^2-p^2\right)$ due to the real-valued artificial pole at $p^2=\Lambda^2$. 

In our calculations, the values of the mixing angle $\theta$ and cutoff $\Lambda$ are constrained by the decay branching ratio ${\rm Br}(X(3872) \to \bar{D}^{*0} D^0) = 34\%$ \cite{ParticleDataGroup:2024cfk}, which also includes the contribution of $D^{*0} \bar{D}^0$ final state~\cite{Belle:2008fma,Belle:2023zxm}: 
\begin{equation} 
\begin{split}
    34\%&=\frac{\Gamma_{X\to \bar{D}^{*0} D^0}}{\Gamma_X}\\
        &=-\frac{1}{\pi\Gamma_X} \int^{(M_X +2\Gamma_X)^2} _{(M_X-2\Gamma_X)^2} \widetilde{\Gamma}_{X\to \bar{D}^{*0} D^0}(\sqrt{s},\theta,\Lambda)\\
        &\quad \times \Theta \left( \sqrt{s}-M_{\bar{D}^0}-M_{D^{*0}} \right) \\
        &\quad \times \text{Im}\left\{\frac{1}{s-M^2_{X}+iM_{X}\Gamma_{X}} \right\}\text{d}s\ , \label{Eq:X3872}
\end{split}
\end{equation}
where $\sqrt{s}$ denotes the total energy of the $\bar{D}^{*0} D^0$ system in the center-of-mass frame, and the energy-dependent width is
\begin{equation} 
    \widetilde{\Gamma}_{X\to \bar{D}^{*0} D^0}(\sqrt{s},\theta,\Lambda) = \frac{|\boldm{p}|}{8\pi s} \frac{1}{3} \sum_{\text{spin}} \left|\CM^{\text{Total}}\right|^2\ .
\end{equation}
Note that it is important to take into account the finite-width effect of $X(3872)$ since it lies very close to the $\bar{D}^{*0} D^0$ threshold. 

The spin effects originating from the charged triangle loop diagram (Fig.~\ref{Fig:Feynman}(b)) are crucial just like in Refs.~\cite{Wang:2022wdm,Wang:2023xua}. Actually we consider here the SDME $\rho_{00}$ of the final $\bar{D}^{*0}$ in the rest frame of $X(3872)$: 
\begin{eqnarray}
    \rho^{\sm{\bar{D}^{*0}}}_{00} (\sqrt{s}) &=& \frac{\int {\rm d} \Omega \sum_{\lambda_1} \CM_{\lambda_1,\lambda_3=0} \CM^*_{\lambda_1,\lambda_3'=0}} {\int{\rm d} \Omega \sum_{\lambda_1,\lambda_3^{''}} |\CM_{\lambda_1,\lambda_3^{''}}|^2}, \label{Eq:Rho00}
\end{eqnarray}
where $\lambda_1$ represents the $z$ component of the total angular momentum of $X(3872)$ and $\lambda_3$, $\lambda_3'$, and $\lambda_3^{''}$ are the helicities of the final $\bar{D}^{*0}$.

\begin{table}[tbp]
    \caption{Coupling constants used in this work.}
    \begin{tabular*}{\columnwidth}{@{\extracolsep\fill}ccccc}
        \hline\hline
        State         & Width                &  Decay             &      Branching      &    $g$  \\[-3pt]
                        & (GeV)                &  channel         &        ratio            &          \\
        \hline
        $X(3872)$     & $1.19\times10^{-3}$  & $\bar{D}^{*0} D^0$ & $3.40\times10^{-1}$ & - \\
        $D^*(2007)^0$ & $5.53\times10^{-5}$  & $D^0 \pi^0$        & $6.47\times10^{-1}$ &  \ 11.88 \\
        $D^*(2010)^+$ & $8.34\times10^{-5}$  & $D^0 \pi^+$        & $6.77\times10^{-1}$ &  \ 16.82 \\ 
        \hline\hline
    \end{tabular*}
    \label{Tab:couplings}
\end{table}

\section{RESULTS AND DISCUSSION}
\label{results}

\begin{figure}[tbp]
    \begin{center}
    \includegraphics[scale=0.4]{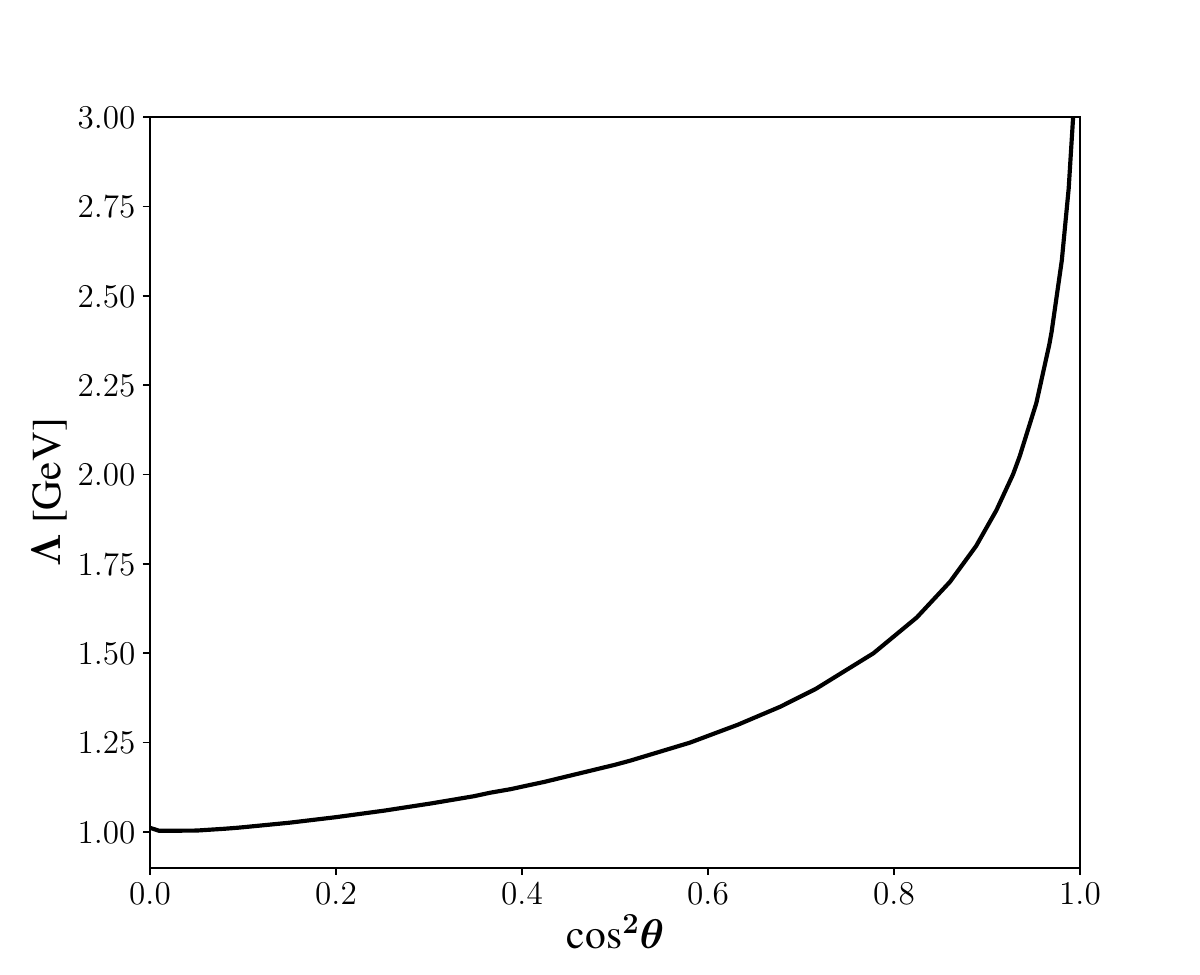}
    \caption{The constraint on the values of $\text{cos}^2\theta$ (neutral portion of $X(3872)$) and $\Lambda$, from the branching ratio of $X\to \bar{D}^{*0} D^0$. Note that $\text{cos}^2\theta$ does NOT physically depend on $\Lambda$. } \label{Fig:theta}
    \end{center}
\end{figure}

By using the package LoopTools~\cite{Hahn:2000jm}, the loop integrals in Eq.~\eqref{Amp:a}, \eqref{Amp:b}, and \eqref{Amp:all} can be calculated numerically. According to Eq.~\eqref{Eq:X3872}, the cutoff $\Lambda$ becomes a function of $\theta$ rather than a free parameter. We first discuss the behavior of $\Lambda(\theta)$. As shown in Fig.~\ref{Fig:theta}, over the interval of the neutral proportion $\cos^2\theta\in[0,1]$, $\Lambda$ is a monotonically increasing function of $\cos^2\theta$ taking the values mainly from the interval $\Lambda\in[1,3]$ GeV, which is reasonable for a GeV-scale physical problem. This behavior is understandable. Notice three facts: first, the partial width $\Gamma_{X\to \bar{D}^{*0} D^0}$ which consists of both the two contributions in Fig.~\ref{Fig:Feynman}, is fixed by Eq.~\eqref{Eq:X3872}; second, the cutoff serves as an overall regulator that adjusts both the two contributions in Fig.~\ref{Fig:Feynman}; third, apart from the proportions, the effective coupling of the neutral channel ($g^{\text{eff}}_n$) is born to be much smaller than that of the charged channel ($g^{\text{eff}}_c$), which is a result of Weinberg's criterion. When $\cos^2\theta$ is small, $\sin\theta$ is relatively large; together with the originally large $g^{\text{eff}}_c$, the charged channel contributes too much to the partial width. Therefore, the cutoff should be smaller to suppress the high-energy contributions in the loop integrals, in order that the total contribution is overall shrunk and Eq.~\eqref{Eq:X3872} is met. Similarly, when $\cos^2\theta$ is large, the charged coupling is suppressed and the total contribution is thus insufficient to satisfy Eq.~\eqref{Eq:X3872}. Then $\Lambda$ increases to compensate that suppression by overall amplifying the loop integrals. It should be emphasized that the physical observable $\theta$ must be experimentally determined and is cutoff-independent. The so-called ``dependence'' in Fig.~\ref{Fig:theta} stems from the certain model parametrization. Anyway, our analysis demonstrates that for any given $\theta$, a suitable cutoff can be selected to reproduce the partial width.

\begin{figure}[tbp]
    \begin{center}
    \includegraphics[scale=0.4]{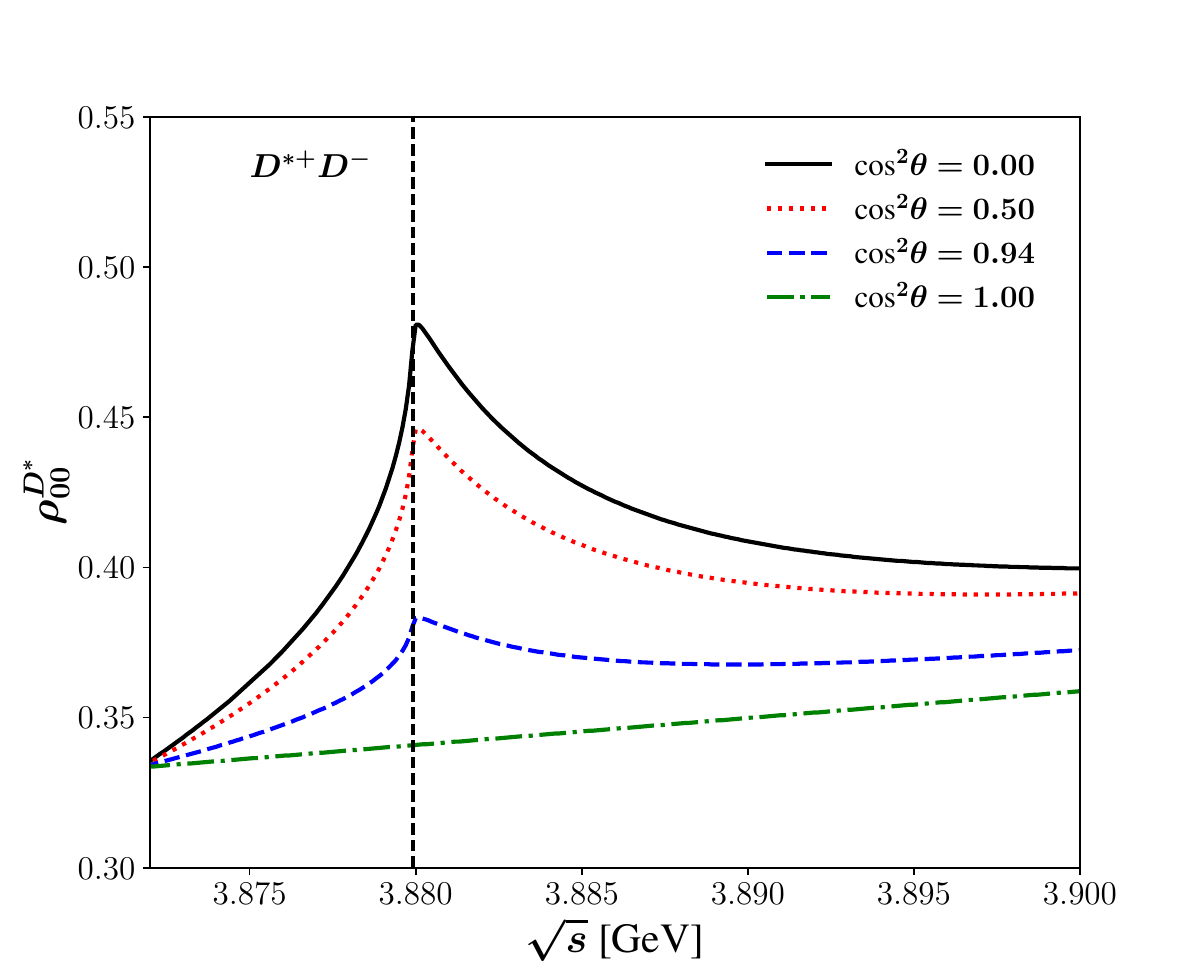}
    \caption{The SDME $\rho^{\sm{D^*}}_{00}$ for the final state $\bar{D}^{*0}$ in the reaction $X(3872) \to \bar{D}^{*0} D$ with different $\text{cos}^2\theta$ values. The vertical line denotes the position of the $D^{*+}D^-$ threshold ($3.880$ GeV). The two critical values: $\cos^2\theta=0.50$ corresponds to a pure isospin singlet $X(3872)$, while $\cos^2\theta=0.94$ corresponds to the point where $g_n=g_c$ exactly holds. } \label{Fig:Rho00_Dstar}
    \end{center}
\end{figure}

Based on Eq.~\eqref{Eq:Rho00}, we present the SDME $\rho^{\sm{D^*}}_{00}$ as the function of the total energy $\sqrt{s}$ in Fig.~\ref{Fig:Rho00_Dstar} when $\text{cos}^2\theta$ takes different values. It can be found that the neutral loop process alone depicted in Fig.~\ref{Fig:Feynman}(a) leads to a nearly straight line shape of $\rho^{\sm{D^*}}_{00}$ within the energy range we concern, as indicated by the green dot-dashed line ($\cos^2\theta=1$). However, when $\cos^2\theta$ deviates from $1$, i.e. the charged configuration enters the process, the curve of $\rho^{\sm{D^*}}_{00}$ displays a distinct structure near the $D^{*+}D^-$ threshold ($\sqrt{s}=3.880$ GeV) due to the threshold singularity produced in Fig.~\ref{Fig:Feynman}(b). Meanwhile, no TS appears in the studied processes. Furthermore, the magnitude of the SDME around the threshold of the charged components depends sensitively on the proportion $\cos^2\theta$. On the other hand, besides the OPE interaction considered in this work, we have also numerically estimated the contributions from vector meson exchanges, such as the exchange of an $\omega$ meson, which turn out to be much smaller than the OPE and do not change the distinctive structures in Fig.~\ref{Fig:Rho00_Dstar}. Consequently, the proportions of the neutral and charged constituents in $X(3872)$ can promisingly be determined if there are experimental data points with enough accuracy near this structure.

\begin{figure}[tbp]
    \begin{center}
    \includegraphics[scale=0.4]{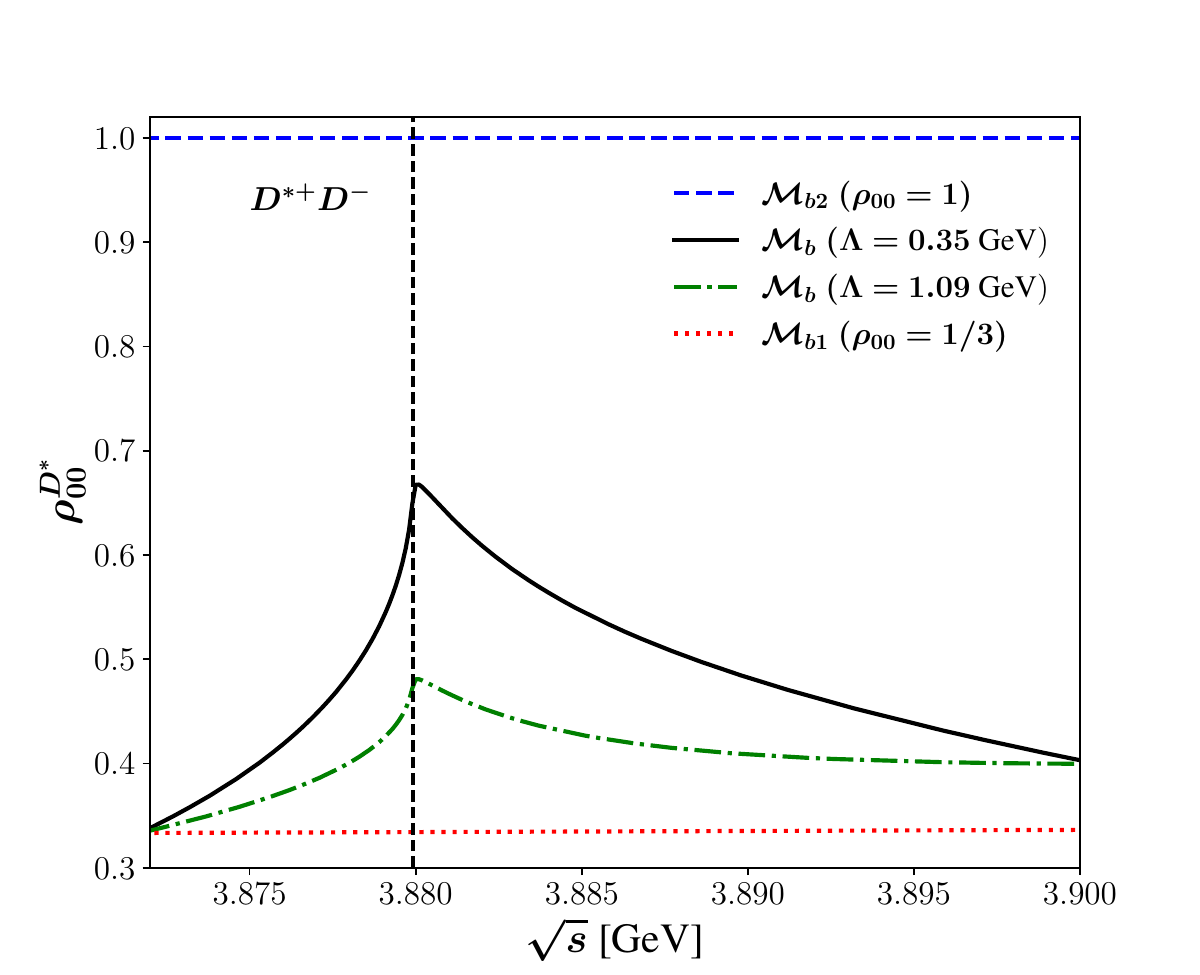}
    \caption{The results of $\rho^{\sm{D^*}}_{00}$ at $\cos^2\theta=0$ with the cutoff from Eq.~\eqref{Eq:X3872} ($1.09$ GeV) versus a smaller value $\Lambda=0.35$ GeV. The constant values $\rho^{\sm{D^*}}_{00}=1/3(1)$ with only $\CM_{b1}(\CM_{b2})$ switched on are also shown here as markers. } \label{Fig:Rho00_12}
    \end{center}
\end{figure}

The behavior of the SDME $\rho^{\sm{D^*}}_{00}$ shown in Fig.~\ref{Fig:Rho00_Dstar} can be further understood, analogous to the discussions in Ref.~\cite{Wang:2022wdm}. Regarding the $D^- + \pi^+ \to \bar{D}^{*0}$ vertex in Fig.~\ref{Fig:Feynman}(b), when the three momenta of all those particles become collinear, angular momentum conservation requires the helicity of the produced $\bar{D}^{*0}$ to be $\lambda_3=0$; then, by definition, $\rho^{\sm{D^*}}_{00}$ takes its ideal maximal value $1$. Such a condition is approximately satisfied when the energy is close to the $D^{*+}D^-$ threshold, namely the momenta of the intermediate $D^{*+}$ and $D^-$ are close to zero and the momenta of $\pi$ and the final state $\bar{D}^{*0}$ are nearly identical. However, in reality $\rho^{\sm{D^*}}_{00}$ does not reach the value $1$, see Fig.~\ref{Fig:Rho00_Dstar}. This is due to the loop momentum $p_{\bar{D}}$ in Eq.~\eqref{Amp:a} which does not stay at one single point in the phase space where the collinear condition holds. In fact the contributions from rest part of the phase space can be controlled by the cutoff parameter. See Fig.~\ref{Fig:Rho00_12}, if we choose a smaller value of $\Lambda$ instead of the one determined by Eq.~\eqref{Eq:X3872}, the contributions beyond the collinear point are suppressed and the actual maximum of $\rho^{\sm{D^*}}_{00}$ becomes closer to $1$. 

To gain more knowledge of the $\rho^{\sm{D^*}}_{00}$ function from a different point of view, we deconstruct the amplitude of the charged configuration. According to Refs.~\cite{Hahn:2000jm,Jing:2019cbw,Du:2022nno}, the amplitude in Eq.~\eqref{Amp:b} can be decomposed into two Lorentz structures
\begin{equation}
    \CM^{\text{Loop}}_{\mu\nu} \equiv C_1 \, g_{\mu\nu} + C_2 \, p_{\sm{\bar{D}^{*0}},\mu} \, p_{\sm{X},\nu},
\end{equation}
where $C_1$ and $C_2$ are energy-dependent scalar functions. Therefore, the charged loop amplitude $\CM_b$ is written as
\begin{eqnarray}
    \CM_b &=& 
    g_b \, \varepsilon^\mu_{X} \varepsilon^{*\nu}_{D^{*0}} \CM^{\text{Loop}}_{b,\mu\nu} 
    \nonumber \\ &=& 
    g_{b}C_1 \, \varepsilon_{X} \cdot \varepsilon^*_{\bar{D}^{*0}} + g_{b}C_2 \, \varepsilon_{X} \cdot p_{\sm{\bar{D}^{*0}}} \, \varepsilon^*_{\bar{D}^{*0}} \cdot p_{\sm{X}} 
    \nonumber \\ &\equiv& 
    \CM_{b1} + \CM_{b2}. \label{Eq:12_wave}
\end{eqnarray}
Note that $\CM_{b1}$ and $\CM_{b2}$ do not correspond to the pure $S$- and $D$-wave contributions, respectively, which has been exhaustively studied in Ref.~\cite{Du:2022nno}. However here a precise partial wave decomposition is unnecessary for our discussions. In fact Eq.~\eqref{Eq:12_wave} is more like a separation of the extra relativistic and basic non-relativistic contributions: under the non-relativistic approximation, $\CM_{b1}$ vanishes while $\CM_{b2}$ remains. If one singly switches $\CM_{b1}$ and $\CM_{b2}$ on, the scalar functions are cancelled in Eq.~\eqref{Eq:Rho00}, leading to trivial constant results $\rho^{\sm{D^*}}_{00}=1/3$ and $\rho^{\sm{D^*}}_{00}=1$, respectively. As shown in Fig.~\ref{Fig:Rho00_12}, the actual SDME lies somewhere in between. Despite the fact that $\rho^{\sm{D^*}}_{00}$ is not a simple superposition of $\CM_{b1}$ and $\CM_{b2}$, it is still expected that the more pronounced $\CM_{b2}$ is, the closer $\rho^{\sm{D^*}}_{00}$ is to $1$. 

\begin{figure}[tbp]
    \begin{center}
    \includegraphics[scale=0.4]{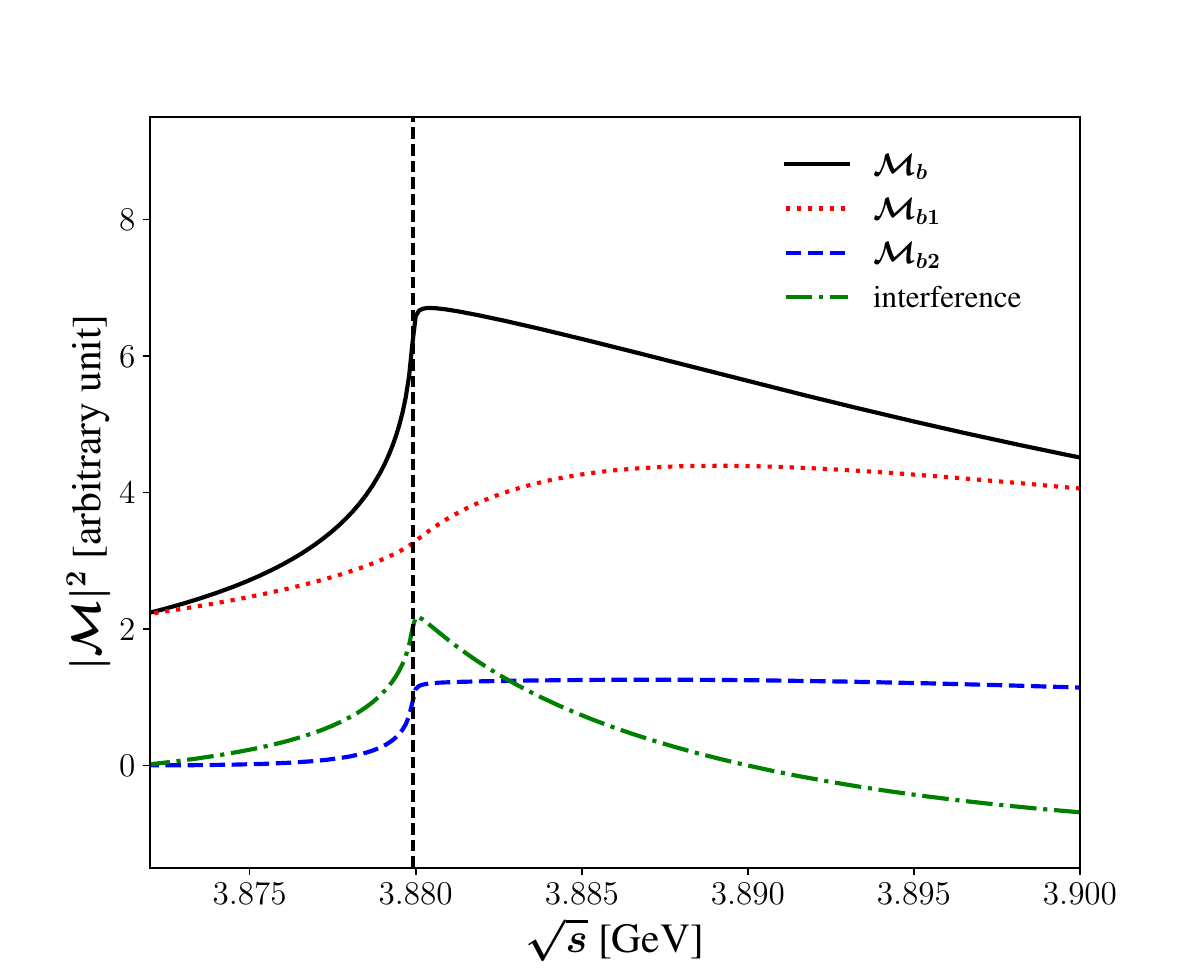}
    \caption{The modulus square of the $\CM_b$ amplitude. The contribution from the interference term between the $\CM_{b1}$ and $\CM_{b2}$ amplitudes is also shown with the green dot-dashed line. } \label{Fig:amp2_12}
    \end{center}
\end{figure}

To see exactly how significant the $\CM_{b2}$ term is as the energy varies, we plot the modulus square of the individual $\CM_{b1}$ and $\CM_{b2}$ terms as well as the whole $\CM_b$ in Fig.~\ref{Fig:amp2_12}. Although the contribution of $\CM_{b1}$ is larger than that of $\CM_{b2}$ all over the energy region, at the $D^{*+}D^-$ threshold there is a sudden pump on the curve of $\CM_{b2}$, making its contribution relatively amplified, and thus the value of $\rho^{\sm{D^*}}_{00}$ is increased. Note that Fig.~\ref{Fig:amp2_12} also indicates the unignorable interference term. The interference effect only affects the terms with $\lambda_3=0$ in Eq.~\eqref{Eq:Rho00}, since $\CM_{b2}$ is only respective to $\lambda_3=0$. At the $D^{*+}D^-$ threshold the interference is constructive, which solely enhances the $\lambda_3=0$ terms in Eq.~\eqref{Eq:Rho00}, consequently enhancing the corresponding peak on the curve of $\rho^{\sm{D^*}}_{00}$. 

In the end, we emphasize that the SDME cannot be replaced by some other observables, e.g. cross sections or invariant mass distributions of certain decay processes, even if they also contain the triangle diagram in Fig.~\ref{Fig:Feynman}(b). The $X(3872)$ state lies very close to the $D^{*+}D^-$ threshold, affecting the line shape in the near-threshold region much by its Breit-Wigner peak. A crucial advantage of the SDME, which many other observables do not possess, is its independence on that Breit-Wigner term -- in case one has to allow for such a term, it appears in both the nominator and denominator of Eq.~\eqref{Eq:Rho00} and is hence cancelled. This prevents the key of our study, i.e. the peak at the $D^{*+}D^-$ threshold caused by the charged configuration, from being submerged in the Breit-Wigner structure. In a word, Fig.~\ref{Fig:Rho00_Dstar} can be regarded as our quantitative prediction of the $\rho^{\sm{D^*}}_{00}$ observable. Especially its value near the threshold: 
\begin{equation}
    \rho^{\sm{D^*}}_{00}(\sqrt{s}=m_{D^{*+}}+m_{D^-})\in (0.34,0.48)
\end{equation}
helps determining the proportions much, once the data is available. Moreover, the SDME proposed here is not the only way to determine the neutral/charged portions of $X(3872)$. For example, comprehensive analyses focusing on the aforementioned $R_X$ values may also provide crucial information. In principle the value of $\cos^2\theta$ determined by other methods can be crosschecked against the value from the SDME, once the relevant data become available.

\section{SUMMARY}  
\label{summary}

In this work, based on the interpretation that $X(3872)$ is a weakly bound $S$-wave molecule of $D^{*} \bar{D}$, we propose a method to determine its neutral ($D^{*0} \bar{D}^{0}$) and charged ($D^{*+} D^{-}$) proportions through the spin density matrix element $\rho^{\sm{D^*}}_{00}$ of the transition $X(3872) \to \bar{D}^{*0} D^{0}$. That transition, under the molecular interpretation, is described by one-loop triangle diagrams with $D^{*0} \bar{D}^{0}$ or $D^{*+} D^{-}$ as intermediate states, exchanging one pion. To evaluate such diagrams, we introduce effective Lagrangians. Specifically, the couplings of $X(3872)$ to $D^{*0} \bar{D}^{0}$ and $D^{*+} D^{-}$ are expressed by the proportions together with the effective couplings from Weinberg's criterion, while the other couplings are pinned down from the corresponding partial decay widths. The only model parameter, namely the cutoff $\Lambda$, turns out to be a monotonically increasing function of the neutral proportion, under the constraint from the $X(3872) \to \bar{D}^{*0} D^{0}$ branching ratio. The numerical result of $\rho^{\sm{D^*}}_{00}$ shows a distinguishing feature: the larger the charged proportion is, the higher and clearer the peak at the $D^{*+} D^{-}$ threshold ($3.880$ GeV) would be, which is thus the key to the determination of such proportions. We further investigate the quantitative behavior of $\rho^{\sm{D^*}}_{00}$, and found that its value at the $D^{*+} D^{-}$ threshold deviates from the ideal value $1$, due to the extra off-shell contributions from the loop integral, which is reasonable. All in all, we predict that $\rho^{\sm{D^*}}_{00}$ ranges between $0.34$ and $0.48$ at the $D^{*+} D^{-}$ threshold as the proportions change, opening a large window for the experimental result. Therefore, we suggest high-precision measurements of the $\rho^{\sm{D^*}}_{00}$ observable, especially data near the $D^{*+} D^{-}$ threshold, by future experiments such as BESIII and Belle.

\begin{acknowledgements}
This work is supported by the National Natural Science Foundation of China under Grants No.~12347155 and No.~12175240, the China Postdoctoral Science Foundation under Grant No.~2024M753172, and the Fundamental Research Funds for the Central Universities.
\end{acknowledgements}

\end{document}